\documentstyle[prd,aps,eqsecnum,epsf,psfig,epsfig]{revtex}
\newcommand {\be}       {\begin{equation}}
\newcommand {\ee}       {\end{equation}}
\newcommand{\Ib}        {I\hspace{-1.55mm}\rule[1.1mm]{1.2mm}{0.15mm}}
\newcommand{\DI}        {\Delta I}
\newcommand{\phidd}     {\ddot{\phi}}
\newcommand{\phid}      {\dot{\phi}}
\newcommand{\npnd}      {\mbox{\boldmath$n_{\perp n_{d}}$}}
\newcommand{\npJ}       {\mbox{\boldmath$n_{\perp J}$}}

\newcommand{\bfv}       {\mbox{\boldmath$\hat{v}$}}
\newcommand{\st}        {\sin \theta}
\newcommand{\ct}        {\cos \theta}
\newcommand{\sst}       {\sin^{2} \theta}
\newcommand{\cst}       {\cos^{2} \theta}
\newcommand{\nd}        {\mbox{\boldmath$n_{d}$}}
\newcommand{\bfT}       {\mbox{\boldmath$T$}}
\newcommand{\J}         {\mbox{\boldmath$J$}}

\newcommand{\bfI}       {\mbox{\boldmath $I$}}
\newcommand{\bfIN}      {\mbox{\boldmath $I_{N}$}}
\newcommand{\dbfIBT}    {\delta \mbox{\boldmath $I_{BT}$}}

\newcommand{\bfdelta}   {\mbox{\boldmath $\delta$}}
\newcommand{\Om}        {\mbox{\boldmath $\Omega$}}
\newcommand{\bfJ}       {\bf{J}}

\newcommand{\Ios}        {I_{0,S}}
\newcommand{\nom}       {\mbox{\boldmath$n_{\Omega}$}}
\newcommand{\DId}       {\Delta I_{d}}
\newcommand{\DIOm}      {\Delta I_{\Omega}}
\newcommand{\eom}       {\epsilon_{\Omega}}
\newcommand{\psid}      {\dot{\psi}}
\newcommand{\that}      {\hat{\theta}}
\newcommand{\ed}        {\epsilon_{d}}
\newcommand{\nJ}        {\mbox{\boldmath$n_{J}$}}

\begin{document}

\tighten

\title{Gravitational Wave Damping of Neutron Star Wobble}

\author{Curt Cutler}
\address{Max-Planck-Institut fuer Gravitationsphysik,
Albert-Einstein-Institut, Am Muehlenberg 1, \\ 
D-14476 Golm bei Potsdam,
Germany; cutler@aei-potsdam.mpg.de}
\author{David Ian Jones}
\address{Faculty of Mathematical Studies, University of Southampton, \\ 
       Highfield, Southampton, United Kingdom; dij@maths.soton.ac.uk \\
       and Department of Physics and Astronomy, University of Wales, \\
       College of Cardiff, P.O.Box 913, Cardiff, United Kingdom}

\date{\today}

\maketitle
\widetext
\begin{abstract}
\hfil
\parbox{5.6in}{ We calculate the effect of gravitational wave (gw)
back-reaction on realistic neutron stars (NS's) undergoing torque-free
precession.  By `realistic' we mean that the NS is treated as a
mostly-fluid body with an elastic crust, as opposed to a rigid body. We
find that gw's damp NS wobble on a timescale 
$\tau_{\theta} \sim 2\times 10^5$ yr $ [10^{-7}/(\DId/I_{0})]^{2} ({\rm kHz}/ \nu_{s})^{4}$, 
where $\nu_s$ is the spin frequency and $\DId$ is the piece of
the NS's inertia tensor that ``follows'' the crust's principal 
axis (as opposed to its spin axis). We give two different
derivations of this result: one based solely on energy and angular momentum
balance, and another obtained by adding the Burke-Thorne radiation reaction
force to the Newtonian equations of motion. This problem was treated long
ago by Bertotti and Anile (1973), but their claimed result is wrong.  When
we convert from their notation to ours, we find that their $\tau_{\theta}$
is too short by a factor $\sim 10^5$ for typical cases of interest,
and even has the wrong sign for $\DId$ negative. We show where their
calculation went astray.  }

\end{abstract} 
\twocolumn
\narrowtext

\section{Introduction}\label{sec:intro}

This paper calculates the effect of gravitational wave (gw) back-reaction
on the torque-free precession, or wobble, of realistic, spinning neutron
stars (NS's). By `realistic' we mean the NS is treated as a mostly-fluid
body with an elastic crust, as opposed to a rigid body. (However we do not
include any superfluid effects in our analysis.)  Freely precessing neutron
stars are a possible source for the laser interferometer gw detectors
(LIGO, VIRGO and GEO under construction, TAMA already operational); it is
the prospect of gravitational wave astronomy that motivated our study.
Also, the first clear observation of free precession in a pulsar signal was
reported very recently \cite{sls00}, making this investigation
all the more timely.

The effect of gw back-reaction on wobbling, axisymmetric {\it rigid} bodies
was first derived 27 years ago, in an impressively early calculation by
Bertotti and Anile \cite{ba73}. They found (correctly) that for rigid
bodies, gw backreaction damps wobble on a timescale (for small wobble angle
$\theta$) $\tau_{\theta}^{\rm rigid} = 1.8 \times 10^6 {\rm \, yr \,}
[10^{-7}/(\Delta I/I_1)]^{2} ({\rm kHz}/\nu_{s})^4 (10^{45} \, {\rm g \,
cm^{2}}/I_{1}$),
%
where $\nu_s$ is the spin frequency and $\Delta I = (I_3-I_1)$ (with $I_1=I_2 \ne I_3)$.

In the same paper, Bertotti and Anile \cite{ba73} went on to calculate
the effect of gw back-reaction on wobble for the more realistic
case of an {\it elastic} NS.  
When cast into our notation, their claimed gw 
timescale is  $5I_{1}c^{5}/[2G (2\pi
\nu_{s})^{4} \DIOm \DId]$ where $\DIOm$ is the asymmetry in the moment of
inertia due to centrifugal forces and $\DId$ is the asymmetry due to some
other mechanism, such as strain in the solid crust. Taking $\DIOm$ to 
be (roughly) the asymmetry expected for a rotating fluid according to $\DIOm/I
\approx 0.3 (\nu_{s}/ \rm kHz)^{2}$, we would then have a damping time of
merely $0.6 {\rm \, yr \,}({\rm kHz}/
\nu_{s})^{6}[10^{-7}/(\DId/I)](10^{45}{\rm g \,
cm^{2}}/I)$. 
Despite the fundamental beauty of this probem and its potential
astrophysical significance, their
remarkable claim--that in realistic NS's, gw's damp wobble with 
amazing efficiency--was apparently little known.
(A citation index search showed that Bertotti and Anile \cite{ba73} 
had been referenced by other authors only four times in the last 27 years.)

We will show that the Bertotti and Anile 
result for elastic NS's is {\it very} wrong, however.
For typical cases of interest, their gw
timescale $\tau_{\theta}$ is too short by a factor $\sim 10^5$ !  Moreover,
their calculation even gives the wrong sign (exponential growth instead of
damping) when $\DId$ is negative.~\footnote{Actually, Bertotti
and Anile \cite{ba73} never claim in words that they find unstable growth
of the wobble angle when $\DId < 0$, but that is what is found if one just
takes their formulae and converts from their notation to ours, as above.
Moreover we
have repeated their (flawed) calculation, including their one crucial
error, and seen that it does lead to a prediction of exponential wobble
growth for $\DId$ negative.  The conversion from their notation to ours is
simply $(\delta_{1}I-\delta_{2}I) (\cos^{2}\gamma-\frac{1}{2}\sin^{2}\gamma)
\rightarrow \DId$ and $\delta_{2}I \rightarrow \DIOm$).}
In contrast, we find that gw's always act to damp the
wobble in realistic NS's, just as for rigid bodies.  
While in Nature the typical case will be 
$\DId$ positive, $\DId < 0$ can also occur in principle.
We call attention to this case not because it is common, but
because it highlights how much our result differs from  Bertotti
and Anile \cite{ba73},
and because, in fact, their implicit prediction of exponential
wobble growth for this case provided our initial impetus 
to look more closely at this problem.

The organization of this paper is as follows. In \S 2 we derive the gw
damping timescale for rigid-body wobble, using the mass
quadrupole expressions for the energy and angular momentum radiated to
infinity.  (This derivation 
is actually Exercise 16.13 in the textbook by Shapiro and
Teukolsky \cite{st83}).  We give another derivation of $\tau_{\theta}$ in
\S 3, this time by adding the Burke-Thorne radiation reaction force
directly to the Newtonian equations of motion.  
This latter approach was how Bertotti and Anile \cite{ba73}
first calculated (correctly) the gw damping time for wobbling, 
{\it rigid} bodies.

In \S 4 we review standard material on the torque-free precession of
elastic bodies, in the absence of viscous terms or gw back-reaction.  In \S
5 we derive the gw damping timescale $\tau_{\theta}$ in the elastic case,
using energy and angular momentum balance.  In \S 6 we give 
a second derivation of $\tau_{\theta}$ in the elastic case, using the
Burke-Thorne radiation reaction force to evolve the elastic body's free
precession. This was also the strategy of Bertotti and Anile \cite{ba73},
and we show where they went wrong. Briefly, they did not realize that in
addition to torquing the NS, the radiation reaction force also perturbs the
NS's shape (in particular, its inertia tensor).  When solving for the
evolution of the wobble angle, we show that the ``perturbed shape'' term in
the equations of motion almost entirely cancels the gw torque term that
they do include. (Of course, by definition there is no ``perturbed shape''
term in the rigid-body case, which is probably why they forgot this term
when adapting that calculation to the elastic case.)  In \S 7 we describe
how to include the effects of a fluid core in the radiation reaction
calculation.  Finally, in \S 8 we conclude by commenting briefly on the
astrophysical implications of our result.  

We will work in cgs units.

\section{Radiation reaction for a rigid body: Energy and angular momentum balance}
\label{sec:rigid}

The derivation of the wobble damping rate for realistic NS's,
using energy and angular momentum balance, is rather similar
to the corresponding derivation for rigid bodies. 
Here we briefly review the solution to the rigid-body problem, as
a warm-up for tackling the realistic case.

Consider an axisymmetric rigid body with principal axes $\hat x_1,\hat
x_2,\hat x_3$ and principal moments of inertia $I_1 = I_2 \ne I_3$.  Let
the body have angular momentum $\J$, misaligned from $\hat x_3$. Define the
wobble angle $\theta$ by $\J \cdot {\bf\hat x_3} = J {\rm cos}\theta$.  It
is a standard result from classical mechanics that (in the absence of
external torques) the body axis $\hat x_3$ precesses around $\J$ with
(inertial frame) precession frequency $\dot{\phi} = J/I_1$, with $\theta$
constant \cite{ll76}.  Together, the pair $(\theta, \dot{\phi})$ completely
specify the free precession (modulo a trivial constant of integration
specifying $\phi$ at $t=0$).  We wish to calculate the evolution of these
two parameters using the \emph{time-averaged} fluxes $(\dot{E}, \dot{J})$.

Straightforward application of the mass quadrupole formalism \cite{mtw73}
gives
\begin{equation}
\label{ianEdot}
\dot{E} = -\frac{2G}{5c^{5}} \dot{\phi}^{6} (\Delta I)^{2} \sin^{2} \theta
(\cos^{2} \theta + 16 \sin^{2} \theta),
\end{equation}
where $\Delta I = I_{3}-I_{1}$, and
\begin{equation}
\label{ianEdotJdot}
\dot{J} = \dot{E}/\dot{\phi} . 
\end{equation}
It follows from differentiation of $\dot{\phi} = J/I_1$ that
\begin{equation}
\label{ianphidd}
\phidd  =  - \frac{2G}{5c^{5}} \frac{\DI^{2}}{I_{1}} 
             \phid^{5}  \sst (16 \sst + \cst).
\end{equation}
To calculate the rate of change of the wobble angle, rearrange
\begin{equation}
\frac{dE}{dt} =   \left.\frac{\partial E}{\partial J}\right|_{\theta}
                  \frac{dJ}{dt}
               +  \left.\frac{\partial E}{\partial \theta}\right|_{J}
                  \frac{d \theta}{dt}
\end{equation}
to give
\begin{equation}
\label{ianthetad}
\dot{\theta} = \frac{\dot{J} 
          [\dot{\phi} - \left.\frac{\partial E}{\partial J}\right|_{\theta}]}
                        {\left.\frac{\partial E}{\partial \theta}\right|_{J}},
\end{equation}
where Eq.~(\ref{ianEdotJdot}) has been used.  The energy of the body is simply its kinetic energy:
\begin{equation}
\label{ianke}
E = \frac{J^{2}}{2I_{1}} \left[1 - \cos^{2} \theta \frac{\Delta I}{I_{3}}\right],
\end{equation}
and so 
\begin{equation}
\label{iandEbdJ}
 \left.\frac{\partial E}{\partial J}\right|_{\theta} =
 \frac{J}{I_{1}} \left[1 - \cos^{2} \theta \frac{\Delta I}{I_{3}}\right], 
\end{equation}
\begin{equation}
\left.\frac{\partial E}{\partial \theta}\right|_{J} =
\frac{J^{2}}{I_{1}} \cos \theta \sin \theta \frac{\Delta I}{I_{3}}.
\end{equation}
This gives
\begin{equation}
\label{ianrigidtd}
\dot{\theta} = -\frac{2G}{5c^{5}} \frac{\DI^{2}}{I_{1}} 
                \dot{\phi}^{4} \cos \theta \sin \theta 
                (16 \sin^{2} \theta + \cos^{2} \theta).
\end{equation}
We can construct timescales on which the spin-down and alignment occur:
\begin{equation}
\label{iantauphid}
\tau^{\rm rigid}_{\dot{\phi}} = -\frac{\dot{\phi}}{\ddot{\phi}} 
                          = \frac{5c^{5}}{2G} \frac{1}{\phid^{4}}
             \frac{I_{1}}{\DI^{2}} \frac{1}{\sst (16\sst + \cst)},
\end{equation}
\begin{equation}
\label{iantautheta}
\tau^{\rm rigid}_{\theta} 
  = -\frac{\sin \theta}{\frac{d}{dt} \sin \theta} 
  =  \frac{5c^{5}}{2G} \frac{1}{\phid^{4}}
             \frac{I_{1}}{\DI^{2}} \frac{1}{\cst (16\sst + \cst)}.
\end{equation}
Radiation reaction causes both $\dot{\phi}$ and $\sin\theta$ to decrease, regardless of whether the body is oblate or prolate.
Note that in the limit of small wobble angle the inertial precession frequency remains almost constant ($\tau^{\rm rigid}_{\phid} \rightarrow \infty$), while $\theta$ decreases exponentially on the timescale
\begin{equation}
\label{iantimescale}
\tau^{\rm rigid}_{\theta \ll 1} =  \frac{5c^{5}}{2G} \frac{1}{\phid^{4}}
             \frac{I_{1}}{\DI^{2}}. 
\end{equation}
Parameterising:
\be
\label{taugrigid}
\tau_{\theta}^{\rm rigid} = 1.8 \times 10^6 {\rm \, yr \,} 
                   \left(\frac{10^{-7}}{\Delta I/I_1}\right)^{2} \! 
                   \left(\frac{\rm kHz}{\nu_{s}}\right)^4 \!
  \left(   \frac{10^{45} \, {\rm g \, cm^{2}}}{I_{1}}\right) \, .
\ee

In the limit of vanishingly small wobble angle the partial derivative
on the lhs of Eq.~(\ref{iandEbdJ}) becomes what we conventionally call the
`spin frequency' $\Omega$ of the body \cite{og69}.  Eq.~(\ref{ianthetad})
then shows that $\dot{\theta}$ is proportional to the difference
between the inertial precession frequency $\dot{\phi}$ and the spin
frequency $\Omega$.  This difference remains finite as $\theta \rightarrow 0$
according to $\phid-\Omega = (\Delta I/I_1)\Omega[1 + {\cal
  O}(\theta^2)]$.  Thus for a prolate body ($\Delta I < 0$), such as
an American football, the body precesses slower than it spins, while
for an oblate body the inertial precession frequency is higher than the spin
frequency.  Since the denominator in (\ref{ianthetad}) is also proportional
to $\Delta I$, the wobble angle decreases regardless of the
sign of this factor.  This viewpoint will be useful when we consider
the radiation reaction problem for an elastic body.

\section{Radiation reaction for rigid bodies: Local force}

We will now re-derive the spin-down and alignment timescales by adding the Burke-Thorne local radiation reaction force to the equations of motion.  

The Burke-Thorne radiation reaction potential at a point $x$ is given by \cite{mtw73}:
\begin{equation}
\label{BTpot}
\Phi^{RR} = \frac{G}{c^{5}} x^{a}x^{b} \frac{d^{5}\Ib_{ab}}{dt^{5}},
\end{equation}
where $\Ib_{ab}$ denotes the trace-reduced quadrupole moment tensor:
\be
\label{trmoi}
\Ib_{ab} = \int_{V} \rho (x_{a}x_{b} - \frac{1}{3}\delta_{ab} x^{2}) \, dV.
\ee
Note that this is related to the moment of inertia tensor according to
\be
\Ib_{ab} = -I_{ab} - \frac{2}{3}\delta_{ab}\int_{V}\rho x^{2} \, dV,
\ee
with the result that
\be
\label{differences}
\Delta I \equiv I_{3} - I_{1} = -(\Ib_{3} - \Ib_{1}). 
\ee
The radiation reaction force (on a particle of unit mass) 
is $F^{RR}_a = -\partial \Phi^{RR}/\partial x^{a}$.  The instantaneous (\emph{not time-averaged}) torque on a body can easily be shown to be 
\be
\label{gentorque}
T^{a} = \frac{2G}{5c^{5}} \epsilon^{abc} \Ib_{bd} \frac{d^{5}\Ib_{dc}}{dt^{5}}.
\ee
Making use of Eq.~(\ref{differences}) it is straightforward to calculate this  torque for the free precessional motion.  We find
\be
\label{torque}
\bfT = \frac{2G}{5c^{5}} \DI^{2} \phid^{5} \st (16 \sst + \cst) \npnd \, ,
\ee
acting always in the plane containing the angular momentum and the symmetry axis $x_{3}$,  and perpendicular to $\nd$, i.e. along the direction of $\npnd$ shown in Fig.~1.  We will refer to this plane as the \emph{reference plane}.
\begin{figure}

\begin{picture}(100,100)(-50,0)

\thicklines

\put(0,0){\vector(0,1){80}}

\put(0,0){\vector(-2,3){40}}



\put(0,80){\line(-3,-2){36.923}}

\put(-12,72){\vector(-3,-2){10}}

\thinlines
\put(-28.823,53,418){\line(-2,3){3.3}}
\put(-33.523,50.285){\line(3,2){4.7}}

\qbezier(-5.547,8.321)(-2.8,10.6)(0,10)



\put(0,83){$\bf{J}$}


\put(-48,60){$\bf{n_{d}}$}

\put(-3.5,11){$\bf{\theta}$}


\put(-40,73){$      \bf{T} = |T| \bf{   n_{\perp n_{d}}    }         $}

\end{picture}
\caption{For the rigid body the gravitational radiation reaction torque $\bf T$ lies in the reference plane.  It acts perpendicular to the symmetry axis, i.e. along the direction of unit vector $ \bf n_{\perp n_{d}}$     }
\end{figure}
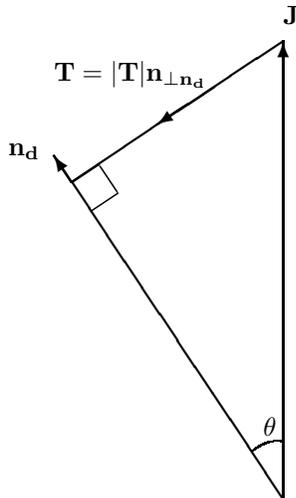
The evolution equations can be calculated without going to the trouble of writing down Euler's equations.  Differentiation of $\phid = J / I_{1}$ gives
\be
\label{phidd}
\phidd = \frac{\dot{J}}{I_{1}}, 
\ee
and so
\be
\label{phiddT}
\phidd = - \frac{T \st}{I_{1}}.
\ee
Define $J_{\perp n_{d}}$ as the component of the angular momentum perpendicular to the symmetry axis.  Then differentiation of the trivial relation
\be
\st = \frac{J_{\perp n_{d}}}{J}        
\ee
leads to 
\be
\label{thetad}
\dot{\theta} = - \frac{ T_{\perp J}} {J} = - \frac{T \ct}{J}, \ee where
$T_{\perp J}$ is the component of the torque perpendicular to $\J$.
Equations (\ref{phiddT}) and (\ref{thetad}) show that the action of the
torque breaks down neatly into two parts.  The component along $\J$ acts to
change the inertial precession frequency $\phid$ while the component
perpendicular to $\J$ acts to change $\theta$.  Substitution of
(\ref{torque}) into (\ref{phiddT}) and (\ref{thetad}) then reproduces the
spin-down and alignment of equations (\ref{ianphidd}) and
(\ref{ianrigidtd}), so the two methods of calculation agree.  As this
torque formulation makes clear (by combining Eqs.~\ref{phiddT} and
{\ref{thetad}), the product $\phid \ct$ remains constant, so that if a body
is set into free precession described by $(\theta_{0}, \phid_{0})$, it tends
to a non-precessing motion about $x_{3}$ with (inertial frame) angular
velocity $\phid = \ct_{0} \,\phid_{0}$.

\section {Torque-free Precession of Elastic Bodies}

We now review the theory of the free precession of an elastic body.
This problem was first addressed in the context of the Earth's own motion.  A rigorous treatment of the methods employed can be found in Munk and MacDonald \cite{mm60}.  The terrestrial analysis was extended to neutron stars by Pines and Shaham \cite{ps72}.  The energy loss due to gravitational waves was considered by Alpar and Pines \cite{ap85}.  

Following the latter authors we will model a star consisting of a centrifugal bulge and a \emph{single} additional deformation bulge.  Alpar and Pines wrote an inertia tensor for the elastic body of the form
\be
\label{moi}
\bfI = \Ios \bfdelta + \DIOm (\nom \nom - 1/3 \bfdelta )
               + \DId (\nd \nd - 1/3 \bfdelta ).
\ee
where $\bfdelta$ is the unit tensor $[1,1,1]$, $\nom$ is the unit vector 
along the star's angular velocity $\bf{\Omega}$, and 
$\nd$ is the unit vector along the body's principal deformation axis 
(explained below).
The $\Ios$  and $\DId$ pieces of $\bfI$ together represent the inertia tensor
for the corresponding {\it non-rotating} star.  The $\DId$ term is just 
the non-spherical piece of this tensor (approximated
as axisymmetric). If the star were a perfect fluid, $\DId$ would
vanish, but in real stars  (and the Earth) $\DId$ is non-zero
due to crustal shear stresses and magnetic fields.
The term $\DIOm$ ($>0$ and $\propto \Omega^2$ for small $\Omega$),
represents the increase in the star's moment of inertia
(compared to the non-rotating case) due to centrifugal forces.
Since the crust of a rotating NS will tend to ``relax'' towards
its oblate shape, having  $\DId >0$ is surely the typical case
in Nature. 
(E.g., if one could slow the Earth down to zero angular velocity
without cracking its crust, it would remain
somewhat oblate: the crust's ``relaxed, zero-strain'' shape is oblate, and
after centrifugal forces are removed, the stresses that build up in the
crust will act to push it back towards that relaxed shape.) 
But a negative $\DId$ is also possible in principle. We say the deformation
bulge aligned with $\nd$ is `oblate' if  $\DId > 0$ and
`prolate' if  $\DId < 0$.

What is a typical magnitude for $\DId$ in real, spinning NS's?  Let us
assume $\DId$ is due primarily to crustal shear stresses (as opposed to
stresses in a hypothetical solid core, extremely strong B-fields, or pinned
superfluid vortices).  Then for a relaxed crust (i.e., a crust whose
reference ellipticity is very close to its actual ellipticity), we have
$\DId = b \DIOm$, where Alpar and Pines~\cite{ap85} estimate $b \sim
10^{-5}$ for a primordial (cold catalyzed) crust.  The maximum value for
$\DId/I$ is therefore of order $\sim 10^{-5}$.  The parameter $b$ (which
arises from inter-nucleon Coulomb forces) scales like the average $Z^2/A$
of the crustal nuclei.  Since crusts of accreted matter (as in LMXB's) have
smaller-Z nuclei \cite{ucb00}, their $b$ factor is correspondingly smaller,
by a factor $\sim 2-3$.  Using $\Delta I_\Omega/I \sim 0.3 (\nu_s/{\rm
kHz})^2$, we would therefore estimate $ \DId/I \sim 10^{-7}$ for a NS with
a relaxed, accreted crust and $\nu_s \sim 300$ Hz, while for the Crab one
would expect $\DId/I \sim 3 \times 10^{-9}$ (again, assuming its crust is
almost relaxed).  For the freely precessing pulsar reported in Stairs et
al.\ \cite{sls00}, where the body-frame precession period is $\sim 2 \times
10^8$ times the rotation period,  Eq.~(\ref{psid}) below (valid
for elastic bodies) yields 
$\DId/I = 5 \times 10^{-9}$.  For $b=10^{-5}$ this corresponds to a
reference oblateness of $5 \times 10^{-4}$.  This is consistent with the
star's crust having solidified when it was spinning at about 40 Hz,
assuming that neither glitches nor plastic flow have modified its shape
since.  (When the effects of crust-core coupling are taken into account, 
giving Eq.~(\ref{psidotcore}),
this initial frequency reduces to 12 Hz.  See Jones \cite{jone00} for a
review of pulsar free precession observations).

Precession occurs when $\nd$ and $\nom$ are not aligned.  Below we
describe the precessional motion when there is no damping.  This
analysis is quite general: it applies to any
star whose inertia tensor is described by Eq.~(\ref{moi}), independent
of what causes the deformation bulge.
In the case of several equally important
sources of deformation along different axes, extra terms must be added
to (\ref{moi}) and the analysis would become more complex.

To proceed it is necessary to use equation (\ref{moi}) to form the
angular momentum $\J$ of the body.  However as we are not modelling a
rigid body, we must take care to allow for relative motion of one part
with respect to another.  Following \cite{mm60} we will write the
velocity of some point in the body as the sum of a rotational velocity
with angular velocity $\Om$ and a small velocity $\bf{u}$ relative to
this rotating frame.  We will call the frame that rotates at $\Om$ the
\emph{body frame} although it is only in the rigid body limit that the
body's shape is fixed with respect to this frame.  In other words the
velocity of some particle making up the body is the sum of the body
frame velocity $\bf \Omega \times r$ at that point $\bf r$ plus the
velocity $\bf u$ of the point relative to the body frame.  Then \be
\label{genj}
J_{a} = I_{ab} \Omega_{b} + h_{a},
\ee
where the possibly time-varying moment of inertia is defined in the usual way:
\be
\label{defnmoi}
I_{ab} = \int_{V} \rho (x_{c}x_{c} \delta_{ab} - x_{a}x_{b} ) \, dV,
\ee
while $h_{a}$ is the angular momentum of the body \emph{relative} to this frame:
\be
\label{hdefn}
h_{a} = \int_{V} \rho \epsilon_{abc} x_{b} u_{c} \, dV.  
\ee
We will neglect the $h_{i}$ term when constructing a free precessional motion, as it can be shown that $h_{i}$ is small in a well-defined sense \cite{jone00}. Therefore we will simply write 
\be
\label{angmomdef}
J_{a} = I_{ab} \Omega_{b}
\ee

Having formulated the problem in this manner it is straightforward to show that the free precession of an elastic body is similar to that of a rigid one.  First write down the angular momentum  using (\ref{moi}) and (\ref{angmomdef}).  Referring all of our tensors to the body frame, with the 3-axis along $\nd$:
\be
\label{angmom}
\J = (\Ios + 2/3\DIOm -1/3\DId ) \Om +  \DId  \Omega{_3} \nd.
\ee
This shows that $\J$, $\Om$ and $\nd$ are coplanar.  As the angular momentum is constant this plane must rotate about $\J$.  As in the rigid body case, we will refer to this as the \emph{reference plane}. See figure \ref{refplane}.  Taking the components of (\ref{angmom}) we obtain:
\be
\label{Jone}
J_{1}  =  (\Ios + 2/3\DIOm -1/3\DId ) \Omega_{1} \equiv   I_{1} \Omega_{1}, 
\ee
\be
\label{Jtwo}
J_{2}  =  (\Ios + 2/3\DIOm -1/3\DId ) \Omega_{2} \equiv   I_{1} \Omega_{2}, 
\ee
\be
\label{Jthree}
J_{3}  =  (\Ios + 2/3\DIOm +2/3\DId ) \Omega_{3} \equiv I_{3} \Omega_{3}.        
\ee
These equations show that despite the triaxiality of $\bfI$ the angular momentum components themselves are structurally equivalent to those of a rigid symmetric top.  The equations of motion of the body (i.e. Euler's equations) involve only the components of $\J$ and $\Om$.  Therefore equations (\ref{Jone})---(\ref{Jthree}) show that the free precession of the triaxial body is formally equivalent to that of a rigid symmetric top.  We can think of the elastic body as having an \emph{effective} moment of inertia tensor $diag[I_{1}, I_{1}, I_{3}]$.  Note that the \emph{effective oblateness} \mbox{$I_{3} - I_{1}$} is equal to $\DId$.

Now introduce standard Euler angles to describe the body's orientation, with the polar axis along $\J$.  Let $\theta$ and $\phi$ denote the polar and azimuthal coordinates of the deformation axis, while $\psi$ represents a rotation about this axis.  We refer to $\theta$ as the \emph{wobble angle}. Taking the ratio of components $J_{1}$ and $J_{3}$ using (\ref{Jone}) and (\ref{Jthree}) at an instant when $\Omega_{2} = 0$   we obtain
\be
\tan \gamma = \frac{I_{3}}{I_{1}} \tan \theta,
\label{angleratio}
\ee
where $\gamma$ denotes the $(\Om ,\nd)$ angle.  See figure \ref{refplane}.
\begin{figure}

\begin{picture}(100,100)(-35,0)


\thicklines
\put(0,0){\vector(0,1){80}}

\put(0,0){\vector(-2,3){30}}

\put(0,83){$\J$}

\put(0,0){\vector(1,4){13}}

\thinlines

\qbezier(-5.547,8.321)(-2.8,10.6)(0,10)

\qbezier(-11.094,16.641)(-3.8,21.7)(4.851,19.403)

\put(13,55){$\bf{\Omega}$}

\put(-30,48){$\nd$}

\put(-30,10){\bf Oblate}

\put(-3.5,11){$\bf{\theta}$}

\put(-4.5,22){$\bf{\gamma}$}

\thicklines
\put(70,0){\vector(0,1){80}}

\put(70,0){\vector(-2,3){30}}


\put(70,83){$\J$}

\put(70,0){\vector(-1,4){13}}

\thinlines

\qbezier(64.453,8.321)(67.2,10.6)(70,10)

\qbezier(58.906,16.641)(63,18.5)(65.6,17.6)

\put(57,55){$\bf{\Omega}$}

\put(40,48){$\nd$}

\put(30,10){\bf Prolate}

\put(63.5,11){$\bf{\theta}$}

\put(60.5,19.5){$\bf{\gamma}$}

\end{picture}
\caption{This shows the reference plane, which contains the deformation axis $ {\bf n_{d}}$, the angular velocity vector $\Om$ and the fixed angular momentum $\bfJ$.  The vectors $n_{d}$ and $\Omega$ rotate around $\bf{J}$ at the \emph{inertial precession frequency} $\dot{\phi}$.  The terms `oblate' and `prolate' refer to the deformation bulge.}
\label{refplane}
\end{figure}
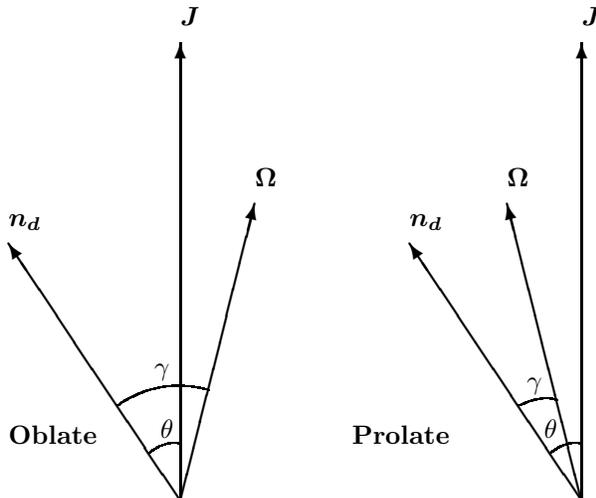
We will label the angle between $\J$ and $\Om$ as $\that$:
\be
\label{thetahat}
\that = \gamma - \theta.
\ee
This angle is much smaller than $\theta$, as can be seen by linearising (\ref{angleratio}) in $\DIOm$ and $\DId$ to give
\be
\label{approxangleratio}
\that = \frac{\DId}{I_{3}} \st \ct.
\ee
Note that according to our conventions, when the deformation bulge is oblate $\DId$ and $\that$ are positive, but when the deformation bulge is prolate $\DId$ and $\that$ are negative.  

We can decompose the angular velocity according to
\be
\label{Omdecompose}
\Om = \phid \nJ + \psid \nd.
\ee
Substituting this into equation (\ref{angmom}) and resolving along $\nJ$ and $\nd$ gives 
\be
\label{phid} 
J = I_{1} \phid,
\ee
\be
\label{psid}
\psid = -\frac{\DId}{I_{1}} \Omega_{3},
\ee
where $J$ denotes the magnitude of the angular momentum.  Note that when $\DIOm = 0$ the above formulae reduce to the familiar rigid body equations.

Thus the motion is simple.  As viewed from the inertial frame the deformation axis rotates at a rate $\phid$ in a cone of half-angle $\theta$ about the angular momentum vector.  This angular velocity is sometimes called the \emph{inertial precession frequency}.  The centrifugal bulge rotates around the angular momentum vector also, but---for oblate deformations---on the opposite side of $\J$, making an angle $\that \equiv \gamma - \theta$ with $\J$. Superimposed upon this is a rotation about the deformation axis at a rate $\psid$, known as the \emph{body frame precession frequency} or sometimes simply the \emph{precession frequency}. This frequency is negative for an oblate distortion and positive for a prolate one.

\section{Radiation reaction for an elastic body: Energy and angular momentum balance}
\label{deriv1}

Here we derive the wobble damping time $\tau_{\theta}$ for elastic bodies,
based on energy and angular momentum balance.  Once fully underway, the
derivation is just a couple lines. But to understand it, it is useful to
carry along a simple, physical model for the deformed crust. (However our
derivation will actually be completely general).  Here is the model: take
some non-rotating, spherical NS, and stretch a rubber band around some
great circle on the crust. We shall refer to this great circle as the NS's
equator. Obviously the effect of the rubber band is to make the NS slightly
prolate (but still axisymmetric). To get an oblate shape, you can instead
imagine sewing compressed springs into the surface of the crust at the
equator. For definiteness, let the potential energy of the band (or
springs) be $V = \frac{1}{2} \epsilon l^2$, where $l$ is its length. So
$\epsilon$ is positive for the rubber band (prolate deformation, $\Delta
I_d < 0$) and negative for the springs (oblate deformation, $\Delta I_d >
0$). Now give the NS angular momentum $\J$ about some axis that is not
quite perpendicular to the equator.  We now have our deformed, wobbling
NS. We consider the equation of state of the star and the value $\epsilon$
to be fixed once and for all, and consider how the energy of the system
(star + band) varies as a function of its total angular momentum $J$ and
the wobble angle $\theta$ (the angle between $\J$ and the perpendicular to
the equator); i.e., we consider $E(J,\theta)$.  We will be concerned with
small wobble angle, so let us expand $E(J,\theta)$ as a Taylor series in
$J$ and $\theta$:

\begin{equation}\label{eq:E_Taylor}
E(J,\theta) = E_0 + \frac{1}{2}B J^2 + \frac{1}{24} C J^4 + \frac{1}{2}F \epsilon \theta^2 J^2 + \cdots 
\end{equation}
Here $E_0$ is defined to be the energy of the (star + band) at zero $J$, and
$B, C$, and $F$ are some expansion coefficients that in principle depend on the physical properties of the (star + band). Fortunately we will soon see that
there  are simple relations between $B, C$, and $F$ and previously-defined
physical parameters, such as $\Delta I_d$.  Our ultimate goal is to obtain the two partial derivatives on the right-hand side of equation (\ref{ianthetad}), where $E$ now denotes the \emph{total} energy.

First, to see that no lower order terms (such as $J, \,\theta J,
\,\theta^2$, or $\theta J^2$ terms) can appear in the expansion
(\ref{eq:E_Taylor}), note that the $J=0$ configuration corresponds to
the minimum of the potential energy of the (star + band)
system. Displacements of the (star + band) are first order in $J^2$,
so changes in the potential energy of (star + band) are ${\cal
O}(J^4)$.  Thus terms in $E(J,\theta)$ that are $\propto J^2$ are
kinetic energy pieces.  These terms with a $J^2$ in them are clearly
just $\frac{1}{2}(I_0^{-1})^{ab}J_aJ_b$, where $I_0^{ab}$ is defined
to be the inertia tensor of the (star + band) at $J=0$.  (Corrections
to the star's $I^{ab}$ first enter the energy at ${\cal O}(J^4)$.)  We
write $I_0^{ab}$ as
\begin{equation}
\label{eq:Iab}
I_0^{ab} = I_{0,S} \delta^{ab} 
+ \Delta I_d \bigr(n_d^an_d^b - \frac{1}{3}\delta^{ab}\bigl),  
\end{equation} 
\noindent where $I_{0,S}$ represents the `spherical part' of $I_0^{ab}$. Then 
\begin{equation}
\label{eq:Iab_inv}
(I_0^{-1})^{ab} =  \frac{1}{I_{0,S}}\left[ \delta^{ab} 
- \left(\frac{\Delta I_d}{I_{0,S}}\right) \bigr(n_d^an_d^b - 
     \frac{1}{3} \delta^{ab}\bigl)\right]
\end{equation}
\noindent where a term of  ${\cal O}(\DId^{2})$ has been neglected.  The kinetic energy part of $E$ is [up to terms of ${\cal O}(\DId^2)$ and ${\cal O}(J^4)$ ]
 \begin{equation}
\label{eq:Ekin}
E_{kin} = \frac{J^{2}}{2I_{0,S}}\left[ 1  -   
 \left(\frac{\Delta I_d}{I_{0,S}}\right) \left(\frac{2}{3} - \theta^2\right) \right]\, ,
\end{equation}
\noindent where we have used the small wobble angle result $J_a n_d^a = J(1 - \frac{1}{2}\theta^2)$. From Eq.~(\ref{eq:Ekin}) we immediately read off the values of $B$ and
$F\epsilon$ in expansion (\ref{eq:E_Taylor}):
\begin{eqnarray}\label{eq:BF}
B =  I_{0,S}^{-1}\bigl[1 -\frac{2}{3}\frac{\Delta I_d}{I_{0,S}} \bigr],
 \nonumber \\ 
F\epsilon = \Delta I_d/(I_{0,S})^2  \, ,
\end{eqnarray}
and obtain the partial derivative
\be
\left.\frac{\partial E}{\partial \theta}\right|_{J} 
       = J^{2} \theta \frac{\DId}{(I_{0,S})^{2}}.
\ee
To compute the partial derivative in the numerator of equation (\ref{ianthetad}) it is sufficient to consider the $\theta \rightarrow 0$ limit \cite{og69} so that
\begin{equation}\label{eq:Omega}
\Omega = \left.\frac{dE}{dJ}\right|_{\theta = 0} = B J  + \frac{1}{6} C J^3,
\end{equation}
where $\Omega$ denotes the spin frequency in the axisymmetric limit.  It is related to the inertial precession frequency by
\be
\Omega = \phid(1 - \DId/I_{0,S}).
\ee
The final physics inputs we need are 
\be
\dot E = - \frac{2G}{5c^{5}} \frac{\DId^{2}}{I_{0,S}} \phid^{6} \theta^{2}
  \label{edot} 
\ee
\be
\dot E = \phid \dot{J}  \label{jdot}  \, .
\ee 
\noindent Eqs.~(\ref{edot}) and ~(\ref{jdot}) follows from the quadrupole formalism in the same way as for the rigid body. 

The necessary pieces have been gathered; substituting into equation (\ref{ianthetad}) gives
\begin{eqnarray}\label{thetadot}
\dot \theta &=&  \frac{\dot J I_{0,S}^{2}}{\theta J^2} \frac{(\phid-\Omega)}{\DId}  \label{t1}\\
&=& -\frac{2G}{5c^5}(\frac{\Delta I_d}{I_{0,S}})^2 I_{0,S} \phid^4 \theta 
  \, .  \label{t2}
\end{eqnarray}
\noindent
This is simply the same spin-down rate as for a rigid body, with the
replacement $(\Delta I/I_1) \rightarrow \epsilon_d$.  This is much
longer than the timescale claimed by Bertotti and Anile \cite{ba73} by a
factor $\DIOm/\DId$, which is typically $\sim 10^5$ or higher. 
 
Finally, the spin-down rate $\phidd$ can be obtained in the same way
as for a rigid body, i.e. by differentiating $\phid = J/I_{1}$ and
using equations (\ref{edot}) and (\ref{jdot}).  Strictly there will also 
be a term in $\dot{I_{1}}$, but this correction will be down by
a factor of order $(\Omega/\Omega_{max})^2$. 
We then obtain the same spin-down as for a
rigid body, again with the replacement $\Delta I \rightarrow \DId$:
\be 
\phidd = - \frac{2G}{5c^{5}} \frac{\DId^{2}}{I_{0,S}} \phid^{5}
\theta^{2}\, .  
\ee

\section{Radiation reaction for an elastic body: Local force} 
\label{deriv2}

We now give a second derivation of the wobble damping rate for an elastic
star, by directly adding the gw radiation reaction force to the
Newtonian equations of motion. Besides being a satisfying consistency check
on the calculation in \S4, by doing this second derivation correctly we can
show where Bertotti and Anile \cite{ba73} went astray.

As was the case for the rigid body, the Burke-Thorne potential will exert a torque on the spinning star.  However, this is not the only effect of the radiation reaction force:  It will  distort the shape of the NS and thus its moment of inertia.  The equation describing the precession is then of the from
\be
\label{eqnofmotion}
\frac{d}{dt} [(\bfIN + \dbfIBT) \Om] = \bfT \, ,
\ee
where $\bfIN$ denotes the Newtonian part of the moment of inertia tensor, $\dbfIBT$ the perturbation in this tensor due to the Burke-Thorne force, and $\bfT$ the Burke-Thorne torque.  It was the $\dbfIBT$ terms that were not included by Bertotti and Anile. Fortunately, these can also be calculated explicitly, as
we show below.

\subsection{Effect of $\Phi^{RR}$ on the NS's Shape}

It is perhaps surprising that one {\it can} explicitly determine
the effect of $\Phi^{RR}$ on the NS's moment of inertia, since
the answer would seem to depend on the NS's mass and the details of its 
equation of state; i.e., one might worry that extra parameters must
be specified even to make the problem well-defined. However the 
point is that (from symmetry arguments) the perturbation $\Delta I_{ij}$
depends only on a {\it single} physical parameter, and this parameter
{\it already} appears in our Newtonian equations of motion.
That parameter is $\Delta I_\Omega/\Omega^2$, the amount of oblateness
caused ``per unit centrifugal force''. 

The point is that {\it both} the centrifugal and
radiation reaction forces have the very special property that
they grow linearly with distance from the
center of the star. This fact, coupled with symmetry arguments,
is enough to determine $\Delta I_{ij}$ in terms of $\Delta I_\Omega/\Omega^2$;
no new physical parameters have to be introduced.

Let $\Phi^\Lambda$ be some external potential of the form 
$\Phi^\Lambda \equiv \Lambda^{ab}x_a x_b$, where 
$\Lambda_{ab}$ is some trace-free tensor. 
Allow this potential to act on the non-rotating (and so spherically symmetric) 
NS; it will induce a perturbation $\Delta I^{ab}$ 
in the NS's inertia tensor. Since the background is spherically symmetric,
the only possibility (to first order in the perturbation) is 
that $\Delta I^{ab} = C \Lambda^{ab}$, where $C$ is some constant  
(i.e, independent of $\Lambda^{ab}$).

We can determine $C$ as follows.  Decompose the centrifugal potential into a spherically symmetric and a trace-free piece:
\begin{equation}\label{centpot}
-\frac{1}{2}\Omega^2 (\delta^{ab}- n^a_\Omega n^b_\Omega)x_a x_b = 
-\frac{1}{3}\Omega^2 x^2  + \Lambda^{ab}_\Omega x_a x_b \, ,
\end{equation}
\noindent where $\Lambda^{ab}_\Omega = \frac{1}{2}\Omega^2 (n^a_\Omega n^b_\Omega - \frac{1}{3} \delta^{ab})$. For small $\Omega$ the perturbed inertia tensor
is  $\Delta I^{ab} = \Delta I_\Omega (n^a_\Omega n^b_\Omega - \frac{1}{3} \delta^{ab})$, so the constant $C$ is just $2 \Delta I_\Omega/\Omega^2$. 

The radiation reaction potential for the freely precessing elastic body can be found by substituting the radiation reaction free motion into equation (\ref{BTpot}) to give:
\be
\Phi^{RR} = - \frac{G}{5c^{5}} x^{a}x^{b} 
             \left[ \DId  \frac{d^{5}}{dt^{5}} (\nd_{a}\nd_{b})
                   +\DIOm  \frac{d^{5}}{dt^{5}} (\nom_{a}\nom_{b}) \right].
\ee
The first term is the potential caused by the motion of the deformation bulge, the second by the centrifugal bulge.  The differentiations of the unit vectors are straightforward.  In the case where $\theta \ll 1$ we can approximate $\nd \approx \nJ + \theta \npJ$ and $\nom \approx \nJ - \that \npJ$, where $\npJ$ is the unit vector in the reference plane which lies perpendicular to $\bfJ$ and points towards $\nd$.  We then find
\be
\Phi^{RR} = -\frac{G}{5c^{5}} \phid^{5} x^{a}x^{b} \phid^{5} 
            [\DId \theta - \DIOm \that] (\bfv_{a}\nJ_{b} + \nJ_{a}\bfv_{b}) \, .
\ee
Here $\bfv$ denotes a unit vector $\nJ \times \npJ$.
Using the prescription described above, these radiation reaction potentials can be converted immediately into perturbations of the moment of inertia tensor:
\be\label{dbfiBT}
\dbfIBT = -\frac{2G}{5c^{5}} \phid^{3} [\DId \DIOm \theta - (\DIOm)^{2}\that]
                                 (\bfv \nJ + \nJ \bfv) \, .
\ee

\subsection{Adding $\Phi^{RR}$ to Equations of Motion}

It now remains to compute the torque $\bfT$ using equation
(\ref{gentorque}).  We obtain four terms, corresponding to the expansion of
the product of $\bf \Ib$ with its fifth time derivative.  Again linearising with respect to $\theta$ we obtain
\be
\label{bfT.eq}
\bfT = \frac{2G}{5c^{5}} \phid^{5} 
      [\DId^{2} \theta - \DId \DIOm \that + \DId \DIOm \theta
       - \DIOm^{2} \that] \npJ.
\ee

Define $\eom \equiv \DIOm/\Ios$ and $\ed \equiv \DId/\Ios$.~\footnote{
Note our definition of $\eom$ differs by a factor $2/3$ from \cite{ap85}, who
set $\eom \equiv \frac{2}{3}\DIOm/\Ios$.}
Then the terms on the rhs of (\ref{bfT.eq}) 
stand in the ratio $\ed/\eom : \ed : 1 : \eom$.
We are now in a position to write down the equation for $d(\bfI_{N} \Om)/dt$. 
Using Eq.~(\ref{dbfiBT}) and the Newtonian motion 
to compute $d[(\dbfIBT) \Om]/dt$, and neglecting terms of order $\theta^2$, we
find that Eq.~(\ref{eqnofmotion}) reduces to 
\be
\label{elastictorque}
  \frac{d}{dt}(\bfI_{N} \Om) 
+ \frac{2G}{5c^{5}} \phid^{5} \theta [\DId \DIOm \theta - (\DIOm)^{2}\that]
          \npJ
\ee
\[
    = \frac{2G}{5c^{5}} \phid^{5} 
      [\DId^{2} \theta - \DId \DIOm \that + \DId \DIOm \theta
       - \DIOm^{2} \that] \npJ.
\]
We see that the last two terms on the rhs are cancelled by terms on
the lhs. This leaves 
\be 
\frac{d}{dt}(\bfI_{N} \Om) =
\frac{2G}{5c^{5}} \phid^{5} [\DId^{2} \theta - \DId \DIOm \that] \npJ.
\ee
The problem has reduced to a rigid-body Newtonian one, with the two torque terms indicated on the right-hand side.  The terms stand in the ratio $1 : \eom$.  In fact, the dominant term is the same as that obtained in the rigid body case with the change $\DI \rightarrow \DId$. 

We therefore find that the alignment rate as calculated using the
local Burke-Thorne formalism agrees with the flux-at-infinity method.
The previous force-based calculation of Bertotti and Anile \cite{ba73}
failed to include the deformation $\dbfIBT$, so that the cancellations
in equation (\ref{elastictorque}) described above did not occur.

Finally, it is easy to show that even when the approximations $\theta
\ll 1$, $\ed \ll 1$ are \emph{not} employed, the effective torques
due to the $\dbfIBT$ terms are still perpendicular to $\bfJ$, so the
spin-down $\phidd$ using this local formalism is necessarily the same
as in the flux-at-infinity method.

\section{Allowance for a liquid core}

We have successfully described the effects of gravitational radiation
reaction on an elastic precessing body.  We will now briefly describe how
to extend this result to the realistic case where the star consists of an
elastic shell (the crust) containing a liquid core.  The Earth itself is
just such a body, and the form of its free precession was considered
long ago.  We will base our treatment on that of Lamb \cite{lamb52}, who
considered a rigid shell containing an incompressible liquid of uniform
density.  To make the problem tractable the motion of the fluid was taken
to be one of uniform vorticity.  We will assume the ellipticity of the
shell, and also the ellipticity of the cavity in which the fluid resides,
are small.  Then the small angle free precession of the combined system can
be found by means of a normal mode analysis of the equations of motion
\cite{lamb52}.

The key points are as follows: The fluid's angular velocity vector does not
significantly participate in the free precession.  Instead it remains
pointing along the system's total angular momentum vector.  The shell
precesses about this axis in a cone of constant half-angle.  The fluid
exerts a force on the shell such that the shell's body frame
precession frequency is increased in magnitude, so that:
\be
\psid = - \phid \frac{\DI}{I_{\rm crust}} 
\ee 
where $\DI$ denotes the difference between the 1 and 3 principal moments of
inertia of the \emph{whole body}, not just the shell.

We now wish to calculate the alignment rate of such a body due to
gravitational radiation reaction.  The averaged energy and angular
momentum fluxes, as well as the instantaneous torque, depend only upon the
orientation of the mass quadrupole of the body, and so are exactly the same
as if the body were rigid, i.e. equations (\ref{ianEdot}),
(\ref{ianEdotJdot}) and (\ref{torque}) apply.  Equations giving the kinetic
energy and angular momentum of the body are given in Lamb \cite{lamb52}.
These can be used to obtain the partial derivatives that appear in equation
(\ref{ianthetad}).  Explicitly, we find 
\be 
\left.\frac{\partial E}{\partial J}\right|_{\theta} = \Omega = \phid +
\psid
\ee
and 
\be
\left.\frac{\partial E}{\partial \theta}\right|_{J} = \phid^{2} \theta \DI.
\ee 
(See Jones~\cite{jone00} for a detailed derivation.)

These lead to an alignment timescale that is $I_{\rm crust}/I$ shorter
than that of equation (\ref{taugrigid}).  This result is confirmed using
the local torque formulation, where 
\be 
\dot{\theta} = -\frac{T_{\perp J}}{I_{\rm crust}}.
\ee 

In the realistic case where both crustal elasticity and core fluidity are
taken into account we can combine the above arguments as described by Smith
and Dahlen \cite{sd81}, i.e. we can take the rigid result and put $I
\rightarrow I_{\rm crust}$ and $\DI \rightarrow \DId$ to give
\begin{eqnarray}
\psid = - \phid \frac{\DId}{I_{\rm crust}} \label{psidotcore} \\
\dot{\theta} = -\frac{2G}{5c^{5}} \frac{\DId^{2}}{I_{\rm crust}} 
                \phid^{4} \theta. \label{thetadotcore}
\end{eqnarray}


\section{Conclusions}

We have shown that the gw damping time for wobble in realistic NS's
has the same form as for rigid bodies, but with the replacement
$\DI^{2}/I_{1} \rightarrow \DId^{2}/I_{\rm crust}$.  This given an
alignment timescale of:
\be
\label{finaltd}
\tau_{\theta} = 1.8 \times 10^5 {\rm \, yr \,} \left(\frac{I_{\rm crust}}
{10^{44}\, {\rm g \, cm^{2}}}\right ) \, \left(   \frac{10^{38} \, {\rm g \, cm^{2}}}{\Delta I_d}\right) \, \left(\frac{\rm kHz}{\nu_{s}}\right)^4 .  
\ee
For the Crab, taking $\ed \sim 3 \times 10^{-9}$, this gives 
$\tau_{\theta} \sim 5 \times 10^{13}$yrs--much longer than
the age of the  universe. For an accreting NS with 
$\ed \sim 10^{-7}$ and $\nu_s \sim 300$ Hz, we
estimate $\tau_{\theta} \sim 2 \times 10^8$ yrs.


Our basic conclusion, then, is that gw backreaction is sufficiently
weak that {\it other} sources of dissipation probably dominate.
Unfortunately, even for
the Earth the dissipation mechanisms are not well understood
\cite{mm60}. Early estimates of Chau and Henriksen \cite{ch71}, which
considered dissipation within the neutron star crust, suggested that wobble
would be damped in around $10^{6}$ free precession periods, i.e. over a
time interval of $10^{6}/(\ed \nu_s)$.
A more recent
study of Alpar and Sauls \cite{as88} argued that the dominant dissipation
mechanism will be due to imperfect coupling between the crust and the
superfluid core.  They estimate that the free precession will be damped in
(at most) $10^{4}$ free precession periods.  In contrast, according to equation
(\ref{finaltd}), the gw damping time is in excess 
of $10^{8}\left(\frac{\rm kHz}{\nu_{s}}\right)^3$ 
free precession periods.  On the basis of
these estimates, it seems likely that internal damping will dominate over
gravitational radiation reaction in \emph{all} neutron stars of interest.
Note however, that while internal dissipation damps 
wobble for \emph{oblate} deformations, we expect that
internal dissipation causes the wobble angle to \emph{increase}
in the prolate ($\DId<0$) case.

A study of the gravitational wave detectability of realistic neutron
stars undergoing free precession, including a discussion of other
astrophysical mechanisms which might affect the evolution of the
motion, will be presented elsewhere (Jones, Schutz and Andersson, in
preparation).

\acknowledgements

We thank N. Andersson and B. F. Schutz for discussions. This research was
supported by NASA grant NAG5-4093 and PPARC grant PPA/G/1998/00606.

\end{document}